\documentclass[12pt]{article}

\usepackage{graphics,amssymb,epsfig,float}
\usepackage[usenames,dvips]{color}
\usepackage{graphicx}% Include figure files
\usepackage{epsfig}% Include figure files
\usepackage{rotating}% Include figure files
\usepackage{dcolumn}% Align table columns on decimal point
\usepackage{bm}% bold math
\usepackage{cite}
\usepackage{amsmath}

\def\lsim{\;\raise0.3ex\hbox{$<$\kern-0.75em\raise-1.1ex\hbox{$\sim$}}\;}
\def\gsim{\;\raise0.3ex\hbox{$>$\kern-0.75em\raise-1.1ex\hbox{$\sim$}}\;}
\def\beq{\begin{equation}}   \def\eeq{\end{equation}}
\def\ba{\begin{array}}       \def\ea{\end{array}}
\def\bea{\begin{eqnarray}}   \def\eea{\end{eqnarray}}
\def\nn{\nonumber}
\def\nl{\newline}

\begin{document}

\begin{titlepage}
\begin{flushright}
LPT Orsay 10-94 \\ 
\end{flushright}

\begin{center}
\vspace{1cm}
{\Large\bf Enhanced di-photon Higgs signal in the Next-to-Minimal
Supersymmetric Standard Model} \\
\vspace{2cm}

{\bf{Ulrich Ellwanger}}
\vspace{1cm}\\
\it  Laboratoire de Physique Th\'eorique, UMR 8627, CNRS and
Universit\'e de Paris--Sud,\\
\it B\^at. 210, 91405 Orsay, France \\

\end{center}

\vspace{1cm}

\begin{abstract}
In the Next-to-Minimal Supersymmetric Standard Model, CP-even Higgs
bosons can have masses in the range of $80-110$~GeV in agreement with
constraints from LEP due to their sizeable singlet component.
Nevertheless their branching ratio into two photons can be more than 10 times
larger than the one of a Standard Model Higgs boson of similar mass
due to a reduced coupling to b quarks. This can lead to a spectacular
enhancement of the Higgs signal rate in the di-photon channel at hadron
colliders by a factor 6. Corresponding scenarios can occur in the
Next-to-Minimal Supersymmetric Standard Model for a relatively low Susy
breaking scale.
\end{abstract}

\end{titlepage}

\section{Introduction}

One of the main motivations of the LHC is the search for the Higgs boson
in the Standard Model (SM) of fundamental interactions or, if realized
in nature, the search for at least one of several Higgs bosons in
corresponding extensions of the SM as by supersymmetry (Susy).

In order to separate possible signals for Higgs bosons from the
background, the experimental groups have to make assumptions on its
production modes, decays and masses. Production modes and decays are
quite well known for the SM, and most of its Susy extensions.
Of course, the experimental groups concentrate on Higgs masses $M_H$
which are not in conflict with unsuccessful searches for Higgs bosons at
LEP, typically $M_H \gsim 110$~GeV \cite{Aad:2009wy} or
$M_H \gsim 115$~GeV \cite{Bayatian:2006zz} both within the SM and the
Minimal Supersymmetric Standard Model (MSSM).

It is well known that, in minimal or general supersymmetric extensions
of the SM, lighter Higgs bosons can exist without contradiction with LEP
bounds. However, usually it is assumed that these are too difficult to
detect at the LHC since, due to LEP bounds, their production rates must
be reduced with respect to the SM. In the present paper we point out
that this reasoning can be wrong: inspite of a somewhat reduced
production rate, Higgs bosons with a mass well below $110$~GeV can be
compatible with LEP bounds \emph{and} be visible at hadron colliders due
to an enhanced branching ratio into the particularly clean di-photon
channel: $H \to \gamma\gamma$. In this channel, the separation of a
Higgs signal from the background is particularly efficient, and it would
be desirable if the present studies for Higgs detection in this channel
would be extended to this lower mass range.

Most importantly, corresponding scenarios can be realized in a Susy
extension of the SM where one of the motivations for Susy (the solution
of the finetuning problem) is solved in a particularly efficient way,
since the Susy breaking scale can be relatively low. This Susy extension
is the Next-to-Minimal Supersymmetric Standard Mo\-del (NMSSM).

The NMSSM is the
simplest supersymmetric (Susy) extension of the SM with a scale
invariant superpotential, i.\,e. where the only dimensionful parameters
are the soft Susy breaking terms. No supersymmetric Higgs mass term
$\mu$ is required, since it is generated dynamically by the vacuum
expectation value (vev) of a gauge singlet superfield $S$. Together with
the neutral components of the two SU(2) doublet Higgs fields $H_u$ and
$H_d$ of the MSSM, one finds
three neutral CP-even Higgs states in this model (see
\cite{Maniatis:2009re,Ellwanger:2009dp} for recent reviews of the
NMSSM). These three states mix in the form of a $3 \times 3$ mass matrix
and, accordingly, the physical eigenstates are superpositions of the
neutral CP-even components of $H_u$, $H_d$ and $S$. (Here and below we
assume no CP-violation in the Higgs sector.) In general, the couplings
of the physical states to gauge bosons, quarks and leptons differ
considerably from the corresponding couplings of a SM Higgs boson.

In the MSSM, the absence of a Higgs signal at LEP \cite{Schael:2006cr}
imposes severe restrictions on the viable parameter space: at tree
level, a SM-like Higgs state (with nearly SM-like couplings to gauge
bosons) would have a mass below $M_Z$, which is by far excluded.
Radiative corrections to the Higgs potential can lift the corresponding
Higgs mass above the LEP limits; however, to this end relatively large
soft Susy breaking terms in the form of stop masses close to 1~TeV are
required, which implies a ``little fine tuning problem'': the natural
value for the negative Higgs mass term $-m^2_{H_u}$ in the Higgs
potential (not to be confused with a physical Higgs mass) is of the
order of the stop masses, which would naturally generate a vev $v_u
\equiv \left<H_u\right>$ of ${\cal{O}}(1$~TeV) (instead of
${\cal{O}}(M_Z)$) unless $-m^2_{H_u}$ is compensated to a large extend
by other terms in the Higgs potential. This requires a tuning of
parameters of ${\cal O}(1\%)$.

This problem is alleviated in the NMSSM: first, the additional Higgs
singlet-doublet coupling $\lambda$ in the superpotential of the NMSSM
allows for a tree level mass of the SM-like Higgs state above $M_Z$,
provided $\tan\beta \equiv v_u/v_d$ is not too large
\cite{Maniatis:2009re,Ellwanger:2009dp}. Second, a Higgs state with a
sizeable singlet component can have a mass well below the lower
LEP-bound of 114.7~GeV on a SM-like Higgs mass 
\cite{Ellwanger:1995ru,Ellwanger:1999ji,Dermisek:2007ah}, without
violating corresponding constraints \cite{Schael:2006cr} on its coupling
to the $Z$ boson. (Here we do not consider regions in parameter space
where unconventional Higgs decays here could be possible.)
In this case, the mass of the next-to-lightest Higgs state of the NMSSM
is naturally above the LEP bound. Most importantly, these NMSSM-specific
scenarios do not require large soft Susy breaking terms.

In the present paper we point out that a Higgs state with a mass in the
80--110~GeV region can have an up to 13 times larger branching ratio
into two photons compared to a SM-like Higgs boson of similar mass, and
a 6 times larger signal rate at hadron colliders. (Around 100~GeV, a
light excess of events in the $b\bar{b}$ final state has been observed
at LEP \cite{Schael:2006cr}.) Inspite of a large singlet component of
such a state, this phenomenon is made possible due to a strong reduction
of its coupling to $b\bar{b}$, and a corresponding reduction of its
total width.

Di-photon Higgs signals at the LHC in the NMSSM have been studied before
in \cite{Moretti:2006sv}. This study concentrated on the possible
detection of several of the Higgs states in the NMSSM, and on scenarios
where the mass of a NMSSM Higgs boson is larger than in the MSSM which
are distinct from the parameter region investigated here. In
\cite{Cao:2011pg}, di-photon Higgs signals of the MSSM, NMSSM and nMSSM
are compared under the assumption of unified soft Susy breaking terms
and the correct dark matter relic density, which seems to exclude again
the parameter region investigated here.

In principle, a reduced coupling of a light Higgs to $b\bar{b}$ is
possible in the MSSM as well \cite{Loinaz:1998ph,Carena:1998gk} if the
lighter physical Higgs state is essentially $H_u$-like (i.\,e. without a
$H_d$-component). There, however, this could only occur for large
$\tan\beta$, larger Higgs masses (due to LEP constraints) and in a
particularly tuned region in parameter space. In the NMSSM, due to the
presence of the singlet, a small $H_d$-component of a light physical
Higgs state is more natural.

Nevertheless, a large signal rate for a Higgs state with a large singlet
component seems paradoxical. In the next section we discuss, after a
brief introduction into the model, the couplings of Higgs states in the
corresponding region of the parameter space of the NMSSM. This allows to
understand the origin of the Higgs decay branching ratios as well as
their production rate in gluon fusion. The last section is dedicated to
conclusions and an outlook.

\section{Properties of light Higgs bosons in the NMSSM}

The NMSSM differs from the MSSM by the presence of the gauge singlet
superfield $S$. The Higgs mass term $\mu H_u H_d$ in the superpotential
$W_{MSSM}$ of the MSSM is replaced by a coupling $\lambda$ of $H_u$ and
$H_d$ to $S$ and a self-coupling $\kappa S^3$, hence the superpotential
$W_{NMSSM}$ is scale invariant:
\bea
W_{NMSSM} &=& \lambda S H_u H_d + \frac{\kappa}{3} S^3
+h_t H_u\cdot Q_3 T^c_R \nn \\
&&+ h_b H_d\cdot Q_3 B^c_R + h_\tau H_d \cdot L_3 \tau^c_R
\label{eq:1}
\eea
where we have confined ourselves to the Yukawa couplings of $H_u$ and
$H_d$ to the quarks and leptons $Q_3, T_R,$ $B_R, L_3$ and $\tau_R$ of the
third generation and, for the first and the last time, the fields denote
superfields. Once $S$ assumes a vev $s$, the first term in $W_{NMSSM}$
generates an effective $\mu$-term
\beq\label{eq:2}
\mu_{eff}=\lambda s\; .
\eeq

Apart from the Yukawa couplings and the standard gauge interactions, the
Lagrangian of the NMSSM contains soft Susy breaking terms in the form of
gaugino masses $M_1$, $M_2$ and $M_3$ for the bino, the winos and the
gluino, respectively, mass terms for all scalars (squarks, sleptons,
Higgs bosons including the singlet $S$) as well as trilinear scalar
self-couplings as $\lambda A_\lambda S H_u H_d$, $\frac{\kappa}{3}
A_\kappa S^3$, $h_t A_t H_u\cdot Q_3 T^c_R$, $h_b A_b H_d\cdot Q_3
B^c_R$ and $h_\tau A_\tau H_d \cdot L_3 \tau^c_R$. It is convenient to
replace the three soft Susy breaking mass terms $m^2_{H_u}$, $m^2_{H_d}$
and $m_S^2$ by $M_Z$, $\tan\beta$ and $\mu_{eff}$ with the help of the
minimization equations of the Higgs potential with respect to $v_u$,
$v_d$ and $s$.

For any choice of the parameters in the Lagrangian, the spectrum of the
model can be computed with help of the code NMSSMTools 
\cite{Ellwanger:2004xm,Ellwanger:2005dv}; we employed the version 2.3.2
which is updated including radiative corrections to the Higgs sector
from \cite{Degrassi:2009yq}. Only points respecting constraints on the
Higgs sector from LEP and from B~physics are retained. (Tevatron
constraints are not relevant for the present region in parameter space.)
The code also allows to compute the various Higgs decay branching ratios
through a suitable generalization of HDECAY \cite{Djouadi:1997yw} to the
NMSSM.

As discussed in the introduction, the Higgs sector of the NMSSM allows
for relatively low values for the soft Susy breaking terms (and
$\mu_{eff}$) which must, however, respect constraints from unsuccessful
direct searches for Susy particles. First results from the LHC
\cite{Khachatryan:2011tk,daCosta:2011hh,daCosta:2011qk} indicate that
gluino and/or u- and d-squark masses are above $\sim 700$~GeV, whereas
the t-squark masses (most relevant for the little finetuning problem in
the MSSM) are not (yet?) constraint. For the specific example discussed
below we make the following choice, motivated to a certain extend by the
renormalization group running from the grand unification scale down to
the weak scale (although the precise values are not important):
%\newline
gaugino masses $M_1$=100~GeV, $M_2$=200~GeV and $M_3$=800~GeV,
%\newline
squark masses of 800~GeV (but 600~GeV for the third generation),
slepton masses of 200~GeV,
%\newline
$A_t = A_b = -300$~GeV, $A_\tau= -200$~GeV, $A_\lambda = 400$~GeV,
$A_\kappa = -100$ GeV, $\mu_{eff}=150$~GeV.
%\newline

For the dimensionless parameters we take $\lambda = 0.634$, $\kappa =
0.3$ and $\tan\beta=3.5$, but we get similar results (see
below) for variations of the latter parameters within
several~\%, and/or somewhat smaller or considerably larger dimensionful
parameters. We did not look for a particularly low fine tuned region in
parameter space, but content ourselves with the relatively low values
for the soft stop mass terms.

For this choice of parameters, the masses of the two lightest physical
CP-even Higgs states are 
\beq\label{eq:3}
M_{H_1} \simeq 98\ \mathrm{GeV}\;,\qquad
M_{H_2} \simeq 122\ \mathrm{GeV}\; .
\eeq
In addition there exist a singlet-like CP-odd Higgs state of mass $\sim
180$~GeV and a nearly degenerate multiplet of CP-even, CP-odd and
charged Higgs states of masses $\sim 500$~GeV; these will play no role
in the following. 

The couplings of the Higgs states depend on their decomposition in the
CP-even weak eigenstates $H_d$, $H_u$ and $S$, which is given by
\bea\label{eq:4}
H_1 &\simeq& -0.008\ H_d - 0.60\ H_u + 0.80\ S\; ,\nn \\
H_2 &\simeq&\ \ \ 0.33\ H_d + 0.75\ H_u + 0.57\ S\; .
\eea

Employing the notation $H_i = S_{i,k}\,H_k$ ($k=d,u,s$), the reduced
tree level couplings (relative to a SM-like Higgs boson) of $H_i$ to $b$
quarks, $t$ quarks and electroweak gauge bosons $V$ are
\bea
\frac{g_{H_i bb}}{g_{H_{SM} bb}} &=& \frac{S_{i,d}}{\cos\beta}\;,\qquad
\frac{g_{H_i tt}}{g_{H_{SM} tt}} = \frac{S_{i,u}}{\sin\beta}\;,\nn \\
\frac{g_{H_i VV}}{g_{H_{SM} VV}} &=& \cos\beta\, S_{i,d} + \sin\beta\,
S_{i,u}\; .
\label{eq:5}
\eea

Clearly, the reduced tree level coupling of $H_1$ to $b$ quarks is very
small for $S_{1,d} \simeq -0.008$. Squark/gluino loops can also
contribute (notably for large $\tan\beta$) to the coupling of $H_1$ to
$b$ quarks via its $S_{1,u}$-component \cite{Hempfling:1993kv,
Hall:1993gn, Carena:1994bv}; in the present case the effective $H_1\,bb$
coupling increases by just about $10\%$ due to this phenomenon. Hence it
is not astonishing that the partial decay width $\Gamma(H_1 \to
b\bar{b})$ is strongly reduced with respect to a SM-like Higgs boson; in
fact the dominant contribution (about 60\%) to $\Gamma(H_1 \to
b\bar{b})$ comes from the dominantly top-quark loop induced $H_1 gg^*$
coupling (where $g$ denotes a gluon) and a subsequent $g^* \to b\bar{b}$
decay. All in all the total width of $H_1$ is smaller than the total
width of a SM-like Higgs boson of similar mass by a factor $\sim
0.04$.

The couplings of Higgs bosons to photons are induced by loop diagrams
dominated by top-quark loops. Hence the coupling of $H_1$ is reduced by
$\frac{g_{H_1 tt}}{g_{H_{SM} tt}} \simeq 0.63$ at first sight, but
contributions from non-SM particles in the loops (mainly stop squarks)
\cite{Djouadi:2005gj} increase $R_\gamma \equiv \frac{g_{H_1
\gamma\gamma}} {g_{H_{SM} \gamma\gamma}}$ to $R_\gamma \simeq 0.72$.
Thus, although the partial width $\Gamma(H_1 \to \gamma\gamma)$
is smaller by a factor $R_\gamma^2 \simeq 0.52$ than the corresponding
width of a SM-like Higgs boson, the branching fraction $BR(H_1 \to
\gamma\gamma)$ is enhanced by a factor $R_\gamma^2 / 0,04 \simeq 12.7$,
the result announced above. (The branching fraction of a SM-like Higgs
boson of similar mass would be $BR(H_{SM} \to \gamma\gamma)\simeq
0.15\%$ \cite{Djouadi:2005gi}.)

Altogether the relevant branching ratios of $H_1$ are given by
\bea\label{eq:6}
BR(H_1 \to gg)\simeq 51\%\;,\qquad &
BR(H_1 \to cc)\simeq 35\%\;,\nn \\
BR(H_1 \to bb)\simeq 9\%\;,\qquad &
BR(H_1 \to WW)\simeq 3\%\;,\nn \\
BR(H_1 \to \gamma\gamma)\simeq 1.9\%\;,\qquad &
BR(H_1 \to \tau\tau)\simeq 0.2\%\;.
\eea

Next we turn to the production cross section for $H_1$, again relative
to the one of a SM-like Higgs boson. The dominant Higgs production
process is via gluon-gluon fusion where, as stated above, the $Hgg$
coupling is induced dominantly by a top-quark loop. Whereas this
contribution is reduced by the factor $\frac{g_{H_1 tt}}{g_{H_{SM} tt}}
\simeq 0.63$ as in the case of the $H_1\gamma\gamma$ coupling, stop
loops \cite{Djouadi:2005gj} and the missing negative contribution from
$b$ quarks (for $M_H \sim 100$~GeV \cite{Djouadi:2005gi}) lead to a
value for $R_g \equiv \frac{g_{H_1 gg}} {g_{H_{SM} gg}}$ of $R_g \simeq
0.69$. Hence the production cross section for $H_1$ via gluon-gluon
fusion is reduced by $R_g^2 \simeq 0.60$; a similar reduction by $\simeq
0.34$ occurs for the less important $H_1$ production process via vector
boson fusion due to the reduced coupling of $H_1$ to electroweak gauge
bosons. All in all the signal rate in $gg \to H_1 \to \gamma\gamma$ is
thus still enhanced by a factor $0.43\times 12.7\sim 6$ relative to a
SM-like Higgs boson of similar mass.

\begin{figure}[hb!]
\begin{center}
%\resizebox{1.0\textwidth}{!}{
%\psfig{file=RvsMH,scale=0.35,angle=-90} 
%}
\includegraphics*[width=0.9\linewidth,height=1.0\linewidth,angle=-90]
{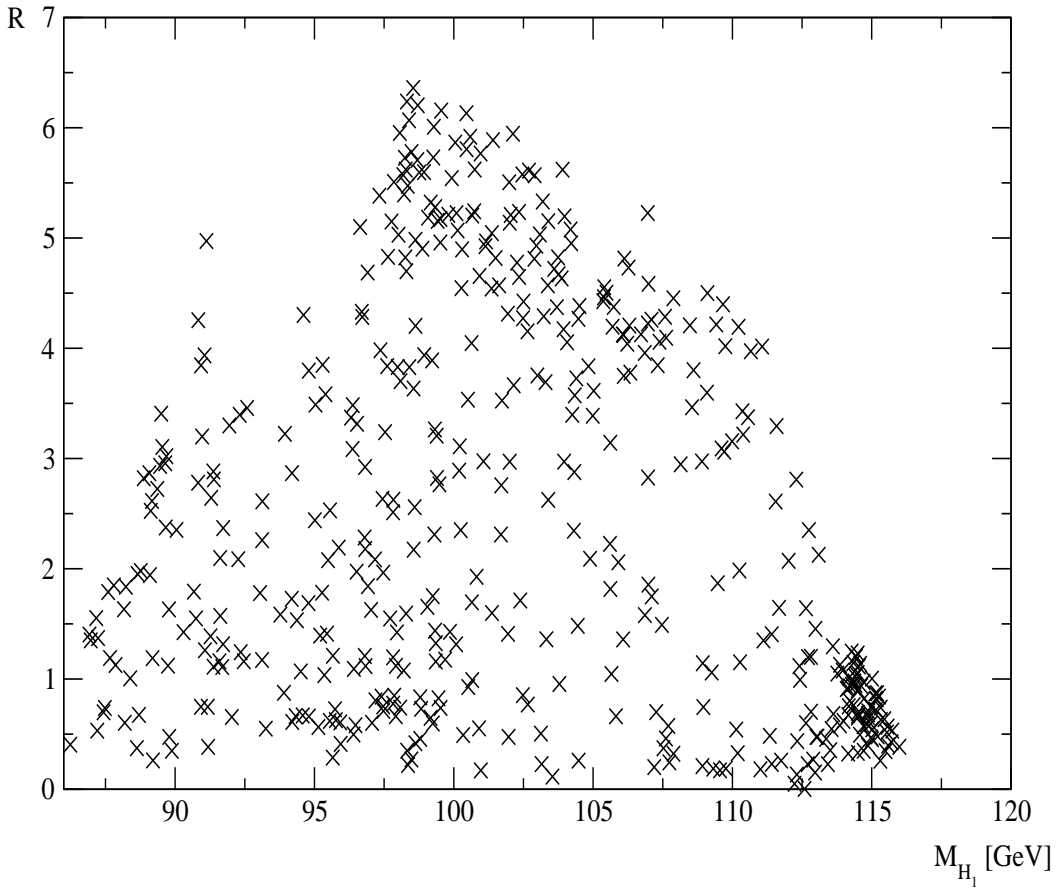}
\caption{The relative signal rate $R=\sigma(gg \to H_1 \to
\gamma\gamma)/\sigma(gg \to H_{SM} \to \gamma\gamma)$ as function of
$M_{H_1}$ for about 500 points in the parameter space of the
NMSSM}
\label{fig:1}
\end{center}
\end{figure}

If we vary the dimensionless parameters in range
$\lambda=0.5-0.7$, $\kappa=0.25-0.35$ and $\tan\beta=3.2-3.5$, the mass
of $H_1$ varies in the range $80-117$~GeV. For parameters outside this
range the mass of $H_1$ can well be below $80$~GeV. Then, however, LEP
constraints imply a very large singlet component of $H_1$ (a reduced
coupling to the Z boson) such that its production rate at the LHC
becomes too small.
Since $S_{1,d}$ can be larger
and $S_{1,u}$ be smaller, the relative signal rate $R=\sigma(gg \to H_1
\to \gamma\gamma)/ \sigma(gg \to H_{SM} \to \gamma\gamma)$ can vary from
$\sim 0$ to $\sim 6.5$, as shown for about 500 points in
Fig.~\ref{fig:1} satisfying LEP and all other phenomenological
constraints. If $R$ is small, the scenario can be similar to the
difficult points discussed in \cite{Ellwanger:2001iw} where a high
luminosity run of the LHC is required in order to detect at least one
Higgs boson of the NMSSM, even though Higgs-to-Higgs decays are not
relevant: due to its reduced couplings, the production rate of $H_2$
will be strongly reduced \emph{without} an enhanced branching ratio into
two photons (of just 0.068\% here). $H_2$ would be most visible in
vector boson fusion and its decay into two tau leptons but, according to
our estimate, more than $\sim 200$~fb$^{-1}$ would be required for its
5~$\sigma$ detection.

The range of $97 - 108$~GeV for $M_{H_1}$, where $R$ can be $\gsim 5$,
overlaps with the range where a light excess (of about 2~$\sigma$
significance) has been observed in the $b\bar{b}$ final state at LEP
\cite{Schael:2006cr}. Here constraints on Higgs bosons with reduced
couplings to the $Z$ boson are relatively weak. Clearly, the
contribution of the state $H_1$ to the LEP signal would be quite small
for a reduced branching fraction into $b\bar{b}$ together with the
reduced coupling to the $Z$ boson. Still, including the $H_1 \to gg^*
\to gbb$ channel, the signal rate at LEP for the points with $R \gsim 5$
in Fig.~\ref{fig:1} can be about 10\% of the one of a SM-like Higgs
boson (possibly enhanced by mis-tagged gluon or charm jets). If we
require at least 5\% for this relative rate, the points with $R \sim 0$
in Fig.~\ref{fig:1} (where $H_1$ is very singlet-like) disappear. Note
that, due to the absence of corresponding contributions from the $gg^*$
channel, the expected excess in the $\tau\tau$ final state is
smaller in agreement with the observations \cite{Schael:2006cr}.

\section{Conclusions and outlook}

We have found that a significant excess of the signal rate in $gg \to
H_1 \to \gamma\gamma$ up to a factor $\sim 6$ with respect to a SM-like
Higgs boson is possible in the NMSSM, remarkably for an
unexpected mass range $M_{H_1} \lsim 110$~GeV. These scenarios are not
far-fetched, since they are possible for a relatively low Susy breaking
scale and motivated, to some extend, by LEP results. (Searches for
``fermiophobic Higgs bosons'' decaying dominantly into two photons had
also been performed at LEP \cite{Rosca:2002me}. However, the upper
limits on $\sigma(h) \times BR(h \to \gamma\gamma)/\sigma(h)_{SM}$ of
$1\%-6\%$ do not exclude the scenarios studied here.)

At present, using data up to 5.4~fb$^{-1}$, the CDF and D0 groups at the
Tevatron exclude to 95\% C.L. a signal in the $\gamma\gamma$
final state if it is about $20-25$ times as large as the one of a SM
Higgs boson for $M_H > 100$~GeV
\cite{:2008it,Aaltonen:2009ga,Peters:2010gd}, hence beyond a possible
signal within the present scenario.

At the LHC with 7 TeV c.m. energy and an integrated luminosity of
1~fb$^{-1}$, the ATLAS \cite{atlas} and CMS \cite{cms} groups expect
95\% C.L. exclusion limits at about 4 times the SM Higgs signal rates
for $M_H > 110-115$~GeV. If the corresponding curves are extrapolated
naively down to $M_H \sim 100$~GeV, the exclusion limits should still be
better than 6 times the SM Higgs signal rate for this mass range.
However, the present results should motivate the experimental groups to
extend their analyses to lower Higgs masses in the $H \to \gamma\gamma$
mode, even if these are seemingly excluded by LEP. At least for 14 TeV
c.m. energy and an integrated luminosity of 30~fb$^{-1}$, signals for a
low mass Higgs boson are well possible in the NMSSM in this channel.


\begin{thebibliography}{99}

\bibitem{Aad:2009wy}
  G.~Aad {\it et al.}  [The ATLAS Collaboration],
  ``Expected Performance of the ATLAS Experiment - Detector, Trigger and
  Physics,''
  arXiv:0901.0512 [hep-ex].

\bibitem{Bayatian:2006zz}
  G.~L.~Bayatian {\it et al.}  [CMS Collaboration],
  ``CMS physics: Technical design report,''

\bibitem{Maniatis:2009re}
  M.~Maniatis,
  Int.\ J.\ Mod.\ Phys.\  A {\bf 25} (2010) 3505
  [arXiv:0906.0777 [hep-ph]].

\bibitem{Ellwanger:2009dp}
  U.~Ellwanger, C.~Hugonie and A.~M.~Teixeira,
  %``The Next-to-Minimal Supersymmetric Standard Model,''
  Phys.\ Rept.\  {\bf 496} (2010) 1\newline
  [arXiv:0910.1785 [hep-ph]].
  
\bibitem{Schael:2006cr}
  S.~Schael {\it et al.} [ALEPH and DELPHI and L3 and OPAL 
  Collaborations and LEP Working Group for Higgs Boson Searches],
  %``Search for neutral MSSM Higgs bosons at LEP,''
  Eur.\ Phys.\ J.\  C {\bf 47} (2006) 547
  [arXiv:hep-ex/0602042].

\bibitem{Ellwanger:1995ru}
  U.~Ellwanger, M.~Rausch de Traubenberg and\nl C.~A.~Savoy,
  %``Higgs phenomenology of the supersymmetric model with a gauge singlet,''
  Z.\ Phys.\  C {\bf 67} (1995) 665
  [arXiv:hep-ph/9502206].

\bibitem{Ellwanger:1999ji}
  U.~Ellwanger and C.~Hugonie,
  %``Masses and couplings of the lightest Higgs bosons in the (M+1)SSM,''
  Eur.\ Phys.\ J.\  C {\bf 25} (2002) 297
  [arXiv:hep-ph/9909260].

\bibitem{Dermisek:2007ah}
  R.~Dermisek and J.~F.~Gunion,
  %``A Comparison of Mixed-Higgs Scenarios In the NMSSM and the MSSM,''
  Phys.\ Rev.\  D {\bf 77} (2008) 015013
  [arXiv:0709.2269 [hep-ph]].

\bibitem{Moretti:2006sv}
  S.~Moretti and S.~Munir,
  Eur.\ Phys.\ J.\  C {\bf 47} (2006) 791
  [arXiv:hep-ph/0603085].
  
\bibitem{Cao:2011pg}
  J.~Cao, Z.~Heng, T.~Liu and J.~M.~Yang,
  ``Di-photon Higgs signal at the LHC: a comparative study for different
  supersymmetric models,''
  arXiv:1103.0631 [hep-ph].

\bibitem{Loinaz:1998ph}
  W.~Loinaz and J.~D.~Wells,
  %``Higgs boson interactions in supersymmetric theories with large
  %tan(beta),''
  Phys.\ Lett.\  B {\bf 445}, 178 (1998)
  [arXiv:hep-ph/9808287].

\bibitem{Carena:1998gk}
  M.~S.~Carena, S.~Mrenna and C.~E.~M.~Wagner,
  %``MSSM Higgs boson phenomenology at the Tevatron collider,''
  Phys.\ Rev.\  D {\bf 60}, 075010 (1999)
  [arXiv:hep-ph/9808312].

\bibitem{Ellwanger:2004xm}
  U.~Ellwanger, J.~F.~Gunion and C.~Hugonie,
  %``NMHDECAY: A Fortran code for the Higgs masses, couplings and decay  widths
  %in the NMSSM,''
  JHEP {\bf 0502} (2005) 066
  [arXiv:hep-ph/0406215].

\bibitem{Ellwanger:2005dv}
  U.~Ellwanger and C.~Hugonie,
  %``NMHDECAY 2.0: An Updated program for sparticle masses, Higgs masses,
  %couplings and decay widths in the NMSSM,''
  Comput.\ Phys.\ Commun.\  {\bf 175} (2006) 290
  [arXiv:hep-ph/0508022].

\bibitem{Degrassi:2009yq}
  G.~Degrassi and P.~Slavich,
  %``On the radiative corrections to the neutral Higgs boson masses in the
  %NMSSM,''
  Nucl.\ Phys.\  B {\bf 825} (2010) 119
  [arXiv:0907.4682 [hep-ph]].

\bibitem{Djouadi:1997yw}
  A.~Djouadi, J.~Kalinowski and M.~Spira,
  %``HDECAY: A program for Higgs boson decays in the standard model and its
  %supersymmetric extension,''
  Comput.\ Phys.\ Commun.\  {\bf 108}, 56 (1998)
  [arXiv:hep-ph/9704448].

\bibitem{Khachatryan:2011tk}
  V.~Khachatryan {\it et al.}  [CMS Collaboration],
  ``Search for Supersymmetry in pp Collisions at 7 TeV in Events with Jets and
  Missing Transverse Energy,''
  arXiv:1101.1628 [hep-ex].

\bibitem{daCosta:2011hh}
  J.~B.~G.~da Costa {\it et al.}  [Atlas Collaboration],
  ``Search for supersymmetry using final states with one lepton, jets, and
  missing transverse momentum with the ATLAS detector in sqrt{s} = 7 TeV pp,''
  arXiv:1102.2357 [hep-ex].

\bibitem{daCosta:2011qk}
  J.~B.~G.~da Costa {\it et al.}  [Atlas Collaboration],
  ``Search for squarks and gluinos using final states with jets and missing
  transverse momentum with the ATLAS detector in sqrt(s) = 7 TeV proton-proton
  collisions,''
  arXiv:1102.5290 [hep-ex].

\bibitem{Hempfling:1993kv}
  R.~Hempfling,
  %``Yukawa coupling unification with supersymmetric threshold corrections,''
  Phys.\ Rev.\  D {\bf 49} (1994) 6168.
 
\bibitem{Hall:1993gn}
  L.~J.~Hall, R.~Rattazzi and U.~Sarid,
  %``The Top quark mass in supersymmetric SO(10) unification,''
  Phys.\ Rev.\  D {\bf 50} (1994) 7048
  [arXiv:hep-ph/9306309].
 
\bibitem{Carena:1994bv}
  M.~S.~Carena, M.~Olechowski, S.~Pokorski and\nl C.~E.~M.~Wagner,
  %``Electroweak symmetry breaking and bottom - top Yukawa unification,''
  Nucl.\ Phys.\  B {\bf 426} (1994) 269\nl
  [arXiv:hep-ph/9402253].
 
\bibitem{Djouadi:2005gj}
  A.~Djouadi,
  %``The Anatomy of electro-weak symmetry breaking. II. The Higgs bosons in the
  %minimal supersymmetric model,''
  Phys.\ Rept.\  {\bf 459} (2008) 1
  [arXiv:hep-\nl ph/0503173].

\bibitem{Djouadi:2005gi}
  A.~Djouadi,
  %``The Anatomy of electro-weak symmetry breaking. I: The Higgs boson in the
  %standard model,''
  Phys.\ Rept.\  {\bf 457} (2008) 1
  [arXiv:hep-\nl ph/0503172].

\bibitem{Ellwanger:2001iw}
  U.~Ellwanger, J.~F.~Gunion and C.~Hugonie,
  ``Establishing a no-lose theorem for NMSSM Higgs boson discovery at the
  LHC'',
  arXiv:hep-ph/0111179.

\bibitem{Rosca:2002me}
  A.~Rosca  [LEP Collaborations],
  %``Fermiophobic Higgs bosons at LEP,''
  arXiv:hep-ex/0212038.

\bibitem{:2008it}
  V.~M.~Abazov {\it et al.}  [D0 Collaboration],
  Phys.\ Rev.\ Lett.\  {\bf 101} (2008) 051801
  [arXiv:0803.1514 [hep-ex]].

\bibitem{Aaltonen:2009ga}
  T.~Aaltonen {\it et al.}  [CDF Collaboration],
  Phys.\ Rev.\ Lett.\  {\bf 103} (2009) 061803
  [arXiv:0905.0413 [hep-ex]].

\bibitem{Peters:2010gd}
  K.~Peters,
  ``Search for the Higgs boson in the gamma gamma final state at the
  Tevatron'',
  arXiv:1009.0859 [hep-ex].

\bibitem{atlas}
  The ATLAS Collaboration, ``ATLAS Sensitivity\nl Prospects for the Higgs
  Boson Production at the LHC Running at 7, 8 or 9 TeV'', ATLAS NOTE
  ATL-PHYS-PUB-2010-015.

\bibitem{cms} 
  The CMS Collaboration, ``The CMS physics reach for searches
  at 7 TeV'', CMS NOTE 2010/008



\end{thebibliography}
\end{document}